\documentstyle[10pt,aas2pp4]{article}
\begin{document}
\newcommand{\etal}{{\it et al.}}
\def\cm2{cm$^2$}
\def\Msun{$M_\odot$}
\def\Mdot{$\dot{M}$}

\begin{flushright}
  {\bf Accepted for Publication in ApJ. Lett. } \\
\end{flushright}

\title{Neutron Star Masses and Radii as Inferred from kilo-Hertz QPOs}

\author{W. Zhang, T. E. Strohmayer$^1$, and J. H. Swank}
\affil{Laboratory for High Energy Astrophysics \\ 
       Goddard Space Flight Center\\
       Greenbelt, Greenbelt, MD 20771\\
       $^1 also$ Universities Space Research Association}

\begin{abstract}

Kilo-Hertz (kHz) Quasi-periodic oscillations (QPOs) have been
discovered in the X-ray fluxes of 8 low-mass X-ray binaries (LMXBs)
with the Rossi X-ray Timing Explorer (RXTE). The characteristics of
these QPOs are remarkably similar from one source to another. In
particular, the highest observed QPO frequencies for 6 of the 8 sources
fall in a very narrow range: 1,066 to 1,171 Hz. This is the more
remarkable when one considers that these sources are thought to
have very different luminosities and magnetic fields, and produce
very different count rates in the RXTE detectors.  Therefore
it is highly unlikely that this near constancy of the highest observed
frequencies is due to some unknown selection effect or instrumental bias. 
In this letter we propose that the highest observed QPO frequency can be
taken as the orbital frequency of the marginally stable orbit. This leads to
the conclusions that the neutron stars in these
LMXBs are inside their marginally stable orbits and have masses in the
vicinity of 2.0\Msun. This mass is consistent with the hypothesis
that these neutron stars were born with about 1.4\Msun and have
been accreting matter at a fraction of the Eddington limit for $10^8$
years.

\end{abstract}


\section{Introduction}

RXTE's large X-ray collection area (7,000 \cm2) and microsecond time
resolution ($2^{-20} s$) combined with its broad telemetry bandwidth
makes it possible for the first time to systematically study fast time
variability in the X-ray fluxes of a large number of galactic sources
(Bradt, Rothschild, \& Swank 1993).  Since its launch on 30 December
1995, kHz QPOs have been discovered in the persistent fluxes of 8
LMXBs: Sco X-1 (van der Klis \etal\ 1996a), 4U 1728-34 (Strohmayer
\etal\ 1996), 4U 1608-52 (Berger \etal\ 1996), 4U 1636-53 (Zhang
\etal\ 1996), 4U 0614+091 (Ford \etal\ 1996), 4U 1735-44 (Wijnands
\etal\ 1996), 4U 1820-30 (Smale, Zhang, \& White 1996), and
GX 5-1 (van der Klis \etal\ 1996d).  In
addition, episodic and nearly coherent oscillations have been discovered 
during several Type-I X-ray bursts from 4U 1728-34 
with a frequency of 363 Hz (Strohmayer \etal\ 1996), 
during one Type-I burst from KS 1731-260 with a frequency of 524 Hz
(Morgan \& Smith 1996), during 3 bursts from the vicinity of 
GRO J1744-28 with a frequency of 589 Hz (Strohmayer, Lee, \& Jahoda 1996),
and during 4 Type-I bursts from 4U 1636-53 with a frequency of
581 Hz (Zhang \etal\ 1997).

Two simultaneously present kHz QPOs have been observed in 6 of the 8
sources, with 4U 1735-44 and 4U 1608-52 showing only one QPO so far.
For Sco X-1 (van der Klis \etal\ 1996b),  the difference in the centroid
frequencies of the two QPOs changes with time (van der Klis
\etal\ 1996b), whereas for 4U 1728-34 and 4U 0614+091, the differences
are constant with time and with count rate (Strohmayer \etal\ 1996 and
Ford \etal\ 1996). In the case of 4U 1728-34, the difference, within
measurement uncertainties, is always 363 Hz, the same as the frequency
of the nearly coherent oscillations observed during several bursts.  In
the case of 4U 1636-53, the difference, which is in the vicinity of 
290 Hz, is quite different from the frequency of the nearly coherent
oscillations during bursts. In the case of 4U 0614+091, the difference
coincides with the centroid frequency of a third QPO which was observed
during one half hour period. Similar information is not available for
4U 1820-30.

These QPOs show remarkably similar characteristics from source to
source.  Their centroid frequencies span a range from a low of 400 Hz
(4U 0614+091, Ford \etal\ 1996) to as high as 1171 Hz (4U 1636-53, van
der Klis \etal\ 1996c). The QPO centroid frequency is correlated with
count rate, reminiscent of the horizontal branch oscillations (HBOs)
observed in Z sources (van der Klis 1995 and references therein), but
with much larger power-law indexes.  An exception to this correlation
is 4U 1608-52 (Berger \etal\ 1996).  The coherence levels of these QPOs
are typically much higher than those of the horizontal branch QPOs of Z
sources. Their Q-values ($\nu/{\Delta\nu}$) can be as high as $10^2$.
Their {\em root-mean-squared} (RMS) amplitudes range from a low at the
threshold of detectability to as high as $12\%$ in the RXTE/PCA
(proportional counter array) band (2--60 keV). In particular, for
every source where this information is available (4U 1728-34, 4U
1608-52, 4U 1636-53), the RMS amplitudes show strong dependence on
energy. For example, in the case of 4U 1636-53, the RMS amplitude at 3
keV is only 4\%, but at 20 keV is as high as 16\% (Zhang \etal\ 1996).  

Three rather different mechanisms have been proposed so far to explain
the existence and characteristics of these QPOs. Klein \etal\ (1996)
have proposed that these QPOs result from turbulence occurring in the
settling mounds of the neutron star polar caps. ``Photon bubbles
oscillations'' (PBO model hereafter) form in the accretion mounds and
transport energy to the surface. Their numerical simulations indicate
QPOs in the kHz range can result with fractional RMS amplitudes 
on the order of 1\%.
Several QPO frequencies have shown up in their
simulations with the highest ones above 2 kHz. Titarchuk and Lapidus
(1996) have proposed a second mechanism in which the kHz QPOs result
from acoustic waves in the boundary region surrounding the
neutron star (hereafter acoustic waves oscillations, or AWO). Miller,
Lamb, and Psaltis (1996) propose a third model in which the QPO with the
higher frequency is the Kepler frequency at the sonic point at which
the radial inflow velocity transitions from subsonic to supersonic. The
position of the sonic point is determined by the radiation forces which
remove the angular momentum of the accreting matter. In this model the
accretion flow is also modulated by the radiation flux from the neutron
star polar caps pulsed at the neutron star spin frequency. This
accretion flow modulation causes the Kepler frequency to beat with the
neutron star spin frequency to produce the QPO with the lower frequency
(hereafter, sonic-point model, or SPM). In this model there is a natural
upper limit to the QPO frequencies, either the Kepler frequency at the 
stellar surface, or the frequency of the marginally stable orbit. 

Due to the semi-quantitative nature of these models at the present and the
insufficient details available from the currently existing data, none
of these models can yet be definitively tested. In this {\em Letter} we
point out the fact that the highest observed frequencies from 6 of
the 8 sources are nearly the same: $\sim 1,100$ Hz. Taking into
account that these sources are believed to have very different
luminosities and magnetic fields, we believe this fact strongly favors
a beat-frequency model, such as the SPM,  where Keplerian orbits play
the most important role in producing the observed variability.

\section{Highest Observed QPO Frequencies}

Table 1 lists the eight LMXBs and their corresponding highest reported
QPO frequency. To our knowledge there has been no systematic effort on the 
part of all the authors to determine the absolutely highest observed 
frequency in each case. Therefore the values should be taken as
what can be determined from the data in a straightforward
analysis. We believe any further dedicated effort to search for the
highest frequencies from these sources will result in qualitatively
similar numbers as in this table.  Even with this caveat in mind, it is
quite striking that the highest observed frequencies of six of the
eight sources fall in a very narrow range: 1066 to 1171 Hz.  Actually,
of the six sources in this range, 5 of them fall in the range of 1130
to 1171 Hz.

We note the two exceptions: 4U 1608-52 and GX 5-1. 4U 1608-52
was observed during the
decline of an outburst (Berger \etal\ 1996). There may be reasons for
this exception. First, the observation may have been too short to have
seen its highest QPO frequency. Its highest QPO frequency may have
occurred during an earlier part of its outburst, thereby missed by RXTE
observations. At any rate the observed frequency is {\it lower} than the
putative upper limit and does not directly contradict any of the following
arguments. GX 5-1 was also observed only for a short time. Most likely
its highest QPO frequency has not been observed.

First, we note that all these reported observations have Nyquist
frequencies much higher than 1100 Hz. Therefore it is impossible for
any problem related to time resolution to have caused all the
highest QPO frequency to be nearly the same for all these sources.
Second, as shown in Table 1, these sources generate
vastly different count rates in the RXTE/PCA detectors, ranging from
over 100,000 counts per second (cps) for Sco X-1 to 400 cps for 4U
0614+091. The counting statistics are vastly different from source to
source. The sensitivity for detecting any QPO is rather different for a
bright source like Sco X-1 from that for a faint source like 4U
0614+091. There appears to be no sensitivity-related limits that
will make all the highest observed QPO frequencies to be the same.
Therefore we conclude that the fact that all these highest frequencies
cluster around 1100 Hz could not have been due to some unknown
effects related to the measurement process, rather it appears to be a
feature common and intrinsic to all these sources.

It is not obvious that there should be any absolute and natural upper
limits on the QPO frequencies in the PBO and AWO models.  Therefore in
this {\em Letter} we will discuss the significance of this near
constancy of the highest observed QPO frequencies in the context of
a beat-frequency model, specifically, the SPM of Miller \etal\ 
In this model, the QPO with the higher centroid
frequency is identified as the Kepler frequency at some radius, and the
nearly coherent oscillations observed during Type-I X-ray bursts (and
the 327 Hz oscillations observed for 4U 0649+091) are identified with
the neutron star spin.  The QPO with the lower centroid frequency
results from beating of these two frequencies. Naturally the radius
where the higher QPO frequency is generated can be identified as the
inner edge of the near-Keplerian flow.

In the magnetospheric beat-frequency model (Alpar \& Shaham 1985)
for the HBOs observed in Z
sources, the inner edge of the accretion disk is at the magnetic
radius, which is determined by the competition of the magnetic stresses
and material stresses, as well as the neutron star
mass. In the most general form, the Kepler frequency ($\nu_K$) at the
magnetic radius can be expressed as
\begin{equation}
   \nu_K \propto \dot{M}^\alpha B^\beta M_{ns}^\gamma,
\end{equation}
where \Mdot\  is the accretion rate, $B$ the magnetic field, and $M_{ns}$
the mass of the neutron star. The power law indices $\alpha, \beta,$
and $\gamma$ depend on the physical conditions relevant for the accretion 
disk. Detailed modeling by Ghosh and Lamb (1991) has shown that 
$\alpha$ range from 0.22 to 2.5, $\beta$ from -1.2 to  -0.76, and
$\gamma$ from -1.0 to 0.07 under various physical assumptions. Barring any
presently unknown correlations between any of the three variables,
in order to have the kind of similar Kepler frequencies at the magnetic radii
for different sources, the accretion rates, and magnetic fields have to be
substantially similar for these sources. For example, in order for
Sco X-1 and 4U 1728-34 to have their Kepler frequencies within 20 Hz
of each other as shown in Table 1, their accretion rates would have 
to be within 10\% of each other. This is contrary to the current
understanding that Sco X-1 is believed to be accreting at or very
near the Eddington limit and that 4U 1728-34 at 10\% or less of the
Eddington limit. Similar arguments apply to the magnetic fields of
these sources. 
Therefore we conclude that it is unlikely that the constancy of
the highest observed QPO frequencies from source to source is
due to the possibility that the edges of the accretion disks, as determined
by the magnetospheres of these sources, are at the same radius.

Miller and Lamb (1993) pointed out that in weakly magnetized neutron
star systems, the accretion disk flow can become significantly
non-Keplerian near the star because radiation forces remove angular
momentum from the accreting matter.  This scenario is used by Miller,
Lamb, and Psaltis (1996) to propose the sonic-point model for the kHz
QPOs. The sonic point is where the radial inflow velocity transitions
from subsonic to supersonic and determined by the luminosity of the
source, which is used by them to explain the kHz QPO frequency and
count rate correlation.

As pointed out by Miller, Lamb, and Psaltis (1996), in the
sonic point model, there is a natural upper limit for the kHz QPO
frequency. It results because either the Keplerian disk extends all the
way to the neutron star surface or that it changes from Keplerian
motion to nearly free fall at the marginally stable orbit. We think
it is unlikely that the highest QPO frequencies are the Kepler
frequencies at the stellar surface, because, if it were so, one would
have to assume the boundary layers of these sources have substantially 
the same size and structure despite the differences in their
magnetic fields and accretion rates. In addition, the sonic point model
depends on the modulated radiation fluxes from the magnetic poles
to generate the lower frequency QPO. If the accretion disk were to
terminate at the stellar surface, one should expect the lower
frequency QPO not to be observed at the same time when the highest
QPO frequency is observed. This is contrary to what has been
observed in 4U 1728-34 (Strohmayer \etal\ 1996) and 4U 
1636-53 (Zhang \etal\ 1997).


Therefore we believe the simplest and most straightforward 
explanation for the near constancy from source to source of
the highest observed QPO frequencies is that they are 
the orbital frequencies of the marginally stable orbits of
their respective neutron stars. If true, this leads to two conclusions.
First, the neutron stars are inside their marginally stable
orbits. Second, the masses of these neutron stars can be
determined from the simple formula (see, e.g., Shapiro \& Teukolsky 1983)
\begin{equation}
   M_{ns}/{M_\odot} = {2198/\nu_{max}.}
\end{equation}
For a typical $\nu_{max}$ of 1100 Hz, one gets $M_{ns}$ = 2 \Msun. 
Here we have ignored the spin of the neutron star, 
which can increase (assuming the disk corotates with the star)
the estimated mass by up to $10-15$\% depending
on the equation  of state and spin rate (see Miller \& Lamb 1996). 

In the next section we discuss the plausibility of these conclusions
in the context of existing literature on neutron star masses and
radii and of the evolution
scenarios of LMXBs.

\section{Discussion}

If our interpretations are correct, we have reached two conclusions:
(1) the neutron stars in LMXBs are inside their marginally stable
orbits; and (2) their masses are of the order of 2\Msun. 

Currently there has been no definitive observational information on 
the radii of
neutron stars. Meszaros and Riffert (1987), synthesizing calculations
of general relativistic effects in the beaming, spectrum, and pulse
properties of accreting neutron stars, and taking account of models for
X-ray pulsars and QPOs observed in X-ray pulsars and LMXBs, argue that a
reasonable value for the neutron star radius in these systems is $4GM/c^2$. 
Note
that the marginally stable orbit has a radius of $6GM/c^2$ for a non-rotating
star. Kluzniak and Wilson (1991) argue that hard X-ray emission from
a weakly magnetic neutron star system can result if one assumes that
the neutron star is smaller than its marginally stable orbit, making it
possible for accreting matter to free-fall to the neutron star surface.
The accreting matter will form a hot, optically thin equatorial
belt with a temperature well above 100 keV. The fact that hard X-rays
have been observed from LMXBs indicate that this model is
plausible (see, e.g., Barret \& Vedrenne 1994).

It is worth noting that nearly all the neutron star models reviewed by
Baym and Pethick (1979) require that a stationary neutron star with mass
above 1.7\Msun be inside its marginally stable orbit. This is even
true for moderately rotating stars with most of the equations of state
(Friedman, Ipser, \& Parker 1986). Therefore we think that in both the
context of theoretical models and observational evidence, it is not
unreasonable to conclude that a neutron star is inside its marginally
stable orbit.

Masses of neutron stars have been measured for a few radio pulsars and
X-ray pulsars (see Nagase 1989 for a review). All these mass measurements
are consistent with 1.4\Msun with Vela X-1 being possibly more massive
by 0.2\Msun. It is generally believed that the neutron stars
in these pulsars are much younger than those in LMXBs. The apparent
difference between our estimate of 2\Msun is not necessarily 
in conflict with any of these measurements.

Estimates for neutron star parameters in LMXBs have come from energy
spectroscopic studies of Type-I X-rays bursts (see Lewin, van Paradijs,
\& Taam 1993 for a review). The results from these studies typically
favor soft equations of state, thereby favoring a lower mass than
1.4\Msun for the neutron stars.  In general, however, these estimates
are plagued by systematics, such as uncertainties of source distances,
relation between effective and color temperatures, the composition of
the accreted matter, etc. Ebisuzaki (1987) in a
particular study of bursts from 4U 1636-536 estimates the neutron star
mass to be 1.8 to 2.0\Msun, in substantial agreement with our estimate
in this {\em Letter}. However, his estimate depends on a
controversial interpretation of  an absorption line observed during
X-ray bursts.

If the neutron stars were born with a mass near 1.4\Msun,
since these LMXBs are believed to have been accreting matter at a
fraction of the Eddington limit for the past $10^8$ years, their
current mass should be about 2.0\Msun, where we have assumed an
accretion energy conversion efficiency of 10\%, and an average
accretion rate of 10\% Eddington.  Therefore we conclude that it is
quite reasonable to have a mass of 2.0\Msun for the neutron stars in
LMXBs, as we have estimated (see, e.g.,  van den Heuvel \& Bitzaraki 1995
for discussioin and references).

In summary, our proposal is based on the available data at the present.
The implicit assumption in making this proposal is that all the neutron
stars in these LMXBs are very similar. They
would all appear to have a mass, within about of 0.2\Msun,
of 2.0\Msun or their mass-radius
combine to give a very similar highest Kepler frequency. Our arguments
appear to strongly support the sonic point model put forth by Miller,
Lamb, and Psaltis (1996).  In this model, there is a clear maximum
frequency for the QPO. It is either the orbital frequency at the
stellar surface or the marginally stable orbit in case the star is
inside the marginally stable orbit. We think the
highest frequencies discussed in this letter are most likely the orbital
frequencies of the marginally stable orbits.

We conclude by pointing out that it now appears possible to measure the
neutron star equation of state by combining the neutron star spin
frequencies ($\nu_{ns}$) measured during X-ray bursts and the mass
estimates ($M_{ns}$) using the highest QPO frequencies. A neutron star
in the process of accreting matter to spin itself up will trace a path
in the $\nu_{ns} - M_{ns}$ plane
which is determined by its equation of state. If we assume that the
neutron stars in LMXBs were born with the same mass and with similar or
slow spin frequencies, their measured $\nu_{ns}$ and $M_{ns}$ will
presumably form a well-defined curve, reflecting that they may have
started accreting at different times and have been accreting with 
different average accretion rates (see, e.g., Lipunov 1992). We realize
that there may be many complicating factors that can muddle this
simplistic picture, nevertheless we think it represents a potentially
very fruitful investigation.  We also point out that it would be of
great interest to study the continuum of the FFT power spectrum to the
highest frequency possible. If our proposal outlined in this paper is
correct, we should expect all these sources will have a feature in
their power spectra at $\sim 1100$ Hz. Searching for this feature or
for the absolute maximum QPO frequency for each source will help verify
our proposal and will quantitatively measure the neutron star masses.

\acknowledgements We wish to thank E. Boldt, F. E. Marshall, and
especially M. C. Miller and F. K. Lamb for many discussions that have
helped us make the arguments in this {\em Letter}.

\begin{table*}
\begin{center}
\bigskip
\caption{Highest observed QPO frequencies from the eight sources. Note
	 that the {\em RXTE/PCA Count Rate} in this table is obtained
	 by scaling the RXTE/ASM light curves of these sources over the
	 period between February and October 1996. The sources are 
	 arranged according to their highest observed QPO frequencies.}
	 \vskip 0.3 in
\begin{tabular}{|lccl|}
\hline
  Source    &  Highest Obs.  & RXTE/PCA Count & Reference \\
	    &  Frequency (Hz)&  Rate (cps)    &         \\
\hline\hline
4U 1636-53  & 1171 & 1,000 to 3,000 &van der Klis \etal\ 1996c.\\
4U 1728-34  & 1150 & 500 to 3,000 & Strohmayer \etal\ 1996. \\
4U 1735-44  & 1149 & 800 to 4,000 & Wijnands \etal\ 1996.\\
4U 0614+091 & 1145 & 300 to 1,200 & Ford \etal\ 1996.\\
Sco X-1     & 1130 & 70,000 to 120,000 & van der Klis \etal\ 1996a. \\
4U 1820-30  & 1066 & 1,000 to 5,000 &Smale \etal\ 1996.\\
4U 1608-52  & 890  & 300 to 3,500 & Berger \etal\ 1996.\\
GX 5-1      & 895  & 7,000 to 14,000 & van der Klis \etal\ 1996d. \\
\hline
\end{tabular}
\end{center}
\end{table*}

\end{document}